\documentclass[magazine]{IEEEtran}
\IEEEoverridecommandlockouts
\usepackage{cite}
\usepackage{amsmath,amssymb,amsfonts}
\usepackage{amssymb}
\usepackage{algpseudocode}
\usepackage{amsfonts}
\usepackage{graphicx}
\usepackage{fancyhdr}
\usepackage{cases}
\usepackage{textcomp}
\usepackage{extarrows}
\usepackage{algorithm,algpseudocode}
\usepackage{multirow}
\usepackage{makecell}
\usepackage{booktabs}
\usepackage{caption}
\usepackage{subcaption}
\usepackage{orcidlink}
\hypersetup{hidelinks}

\usepackage{multirow,tabularx}
\usepackage{mathtools}
\usepackage{xcolor}
\usepackage[english]{babel}
\usepackage{amsthm}
\usepackage[left]{lineno}

\IEEEaftertitletext{\vspace{-12mm}}

\usepackage{caption}
\def\BibTeX{{\rm B\kern-.05em{\sc i\kern-.025em b}\kern-.08em
    T\kern-.1667em\lower.7ex\hbox{E}\kern-.125emX}}
    
\begin{document}
\bstctlcite{BSTcontrol}
\title{Bridging Physical and Digital Worlds: Embodied Large AI for Future Wireless Systems}
\author{
	\IEEEauthorblockN{
	Xinquan Wang$^{\orcidlink{0009-0005-9986-7054}}$,
	Fenghao Zhu$^{\orcidlink{0009-0006-5585-7302}}$,
	Zhaohui Yang$^{\orcidlink{0000-0002-4475-589X}}$, 
	Chongwen Huang$^{\orcidlink{0000-0001-8398-8437}}$, 
	Xiaoming Chen$^{\orcidlink{0000-0002-1818-2135}}$,\\
	Zhaoyang Zhang$^{\orcidlink{0000-0003-2346-6228}}$,
	Sami Muhaidat$^{\orcidlink{0000-0003-4649-9399}}$, and
	M\'{e}rouane~Debbah$^{\orcidlink{0000-0001-8941-8080}}$,~\IEEEmembership{Fellow,~IEEE}}
\thanks{
X. Wang, F. Zhu, Z. Yang, C. Huang, X. Chen and Z. Zhang are with College of Information Science and Electronic Engineering, Zhejiang University, Hangzhou 310027, China. (E-mails:
\href{mailto:wangxinquan@zju.edu.cn}{wangxinquan},
\href{mailto:zjuzfh@zju.edu.cn}{zjuzfh},
\href{mailto:yang_zhaohui@zju.edu.cn}{yang\_zhaohui},
\href{mailto:chongwenhuang@zju.edu.cn}{chongwenhuang},
\href{mailto:chen\_xiaoming@zju.edu.cn}{chen\_xiaoming},
\href{mailto:zhzy@zju.edu.cn}{zhzy}\}@zju.edu.cn).

S. Muhaidat is with Computer and Communication Engineering, Khalifa University, Abu Dhabi 127788, UAE (E-mail: \href{mailto:sami.muhaidat@ku.ac.ae}{sami.muhaidat@ku.ac.ae}).

M. Debbah is with KU 6G Research Center, Department of Computer and Information Engineering, Khalifa University, Abu Dhabi 127788, UAE (E-mail: \href{mailto:merouane.debbah@ku.ac.ae}{merouane.debbah@ku.ac.ae}).}
}
\maketitle

\pagestyle{empty} 
\thispagestyle{empty} 
\begin{abstract}
Large artificial intelligence (AI) models offer revolutionary potential for future wireless systems, promising unprecedented capabilities in network optimization and performance. However, current paradigms largely overlook crucial physical interactions. This oversight means they primarily rely on offline datasets, leading to difficulties in handling real-time wireless dynamics and non-stationary environments. Furthermore, these models often lack the capability for active environmental probing. This paper proposes a fundamental paradigm shift towards wireless embodied large AI (WELAI), moving from passive observation to active embodiment. We first identify key challenges faced by existing models, then we explore the design principles and system structure of WELAI. Besides, we outline prospective applications in next-generation wireless. Finally, through an illustrative case study, we demonstrate the effectiveness of WELAI and point out promising research directions for realizing adaptive, robust, and autonomous wireless systems.
\end{abstract}

\section{Introduction}\label{sec:intro}
Next-generation wireless systems and large artificial intelligence (AI) models are increasingly converging, creating a powerful synergy that reshapes the technological landscape and drives mutual advancement. The integration of these wireless large AI models (WLAMs), often built upon powerful vision or large language models (LLMs), has shown promise in network optimization and control \cite{zhou2024large}. Large AI models enhance wireless systems by processing vast data, learning complex patterns for tasks and making decisions \cite{zengWCM}.
\par
Despite these advances, a fundamental bottleneck severely limits the effectiveness of current WLAM paradigms: they operate in isolation and treat intelligence as a disembodied information processing task, lacking a systematic framework to bridge the fundamental gap between their semantic understanding and the dynamic physical interaction of the wireless environment.
First, they lack a systematic framework for deep causal understanding of the environment, often relying on statistical correlations that fail in dynamic conditions. Second, they cannot bridge the gap between abstract knowledge and meaningful physical interaction. For example, a model might recognize a vehicle in an image but has no native ability to understand how that vehicle physically attenuates a signal, because no mechanism exists to ground its knowledge in physical cause and effect.
\par
Recently, the emergence of agentic AI has enabled models to plan, infer, and act towards long-term goals \cite{agenticZRC}. While promising, most current agentic AI applications remain confined to digital worlds (e.g., programming, web navigation), where physical feedback is limited. In wireless systems, where the AI agent operates under strict physical constraints, these limitations become even more pronounced. Specifically, current AI agents cannot effectively understand or interact with wireless environments because their plans are detached from real-world physical feedback. While digital twins provide valuable simulated environments, without direct physical grounding, agents remain limited in validating or refining their strategies for complex, dynamic real-world conditions. As a result, even the most advanced disembodied AI operates in isolation, possessing immense computational potential but lacking the direct physical perception and action capabilities required to shape its environment effectively.
\par
To bridge the AI-physical gap in wireless systems, we introduce the wireless embodied large AI (WELAI) paradigm. WELAI represents a paradigm shift towards active embodiment, moving beyond passive observation to enable continuous, closed-loop interaction between perception and action within the physical environment \cite{TETCIembodied, chenVTM}. This approach is inspired by the success of embodied AI in robotics and autonomous driving, where learning emerges through real-time adaptation to external dynamics. Specifically, a WELAI agent is itself an intelligent physical entity (such as a vehicle or a base station) that is designed not merely to analyze data, but to actively sense the current state, take physically meaningful actions (e.g., adjusting beamforming) to intervene in, and continuously learn from the physical wireless environment through feedback. This enables adaptive, robust, and autonomous wireless systems capable of optimizing network performance and achieving long-horizon operational goals.
\par
The remainder of this article is organized as follows: Section \ref{sec:limit} outlines the challenges that current paradigms face. Section \ref{sec:principles} declares the principles of the WELAI paradigm. Section \ref{sec:design} explores the architecture of WELAI systems. 
Section \ref{sec:case} presents a case study validating the effectiveness of WELAI paradigm, and discussion the potential wider application of WELAI. Potential future research directions are provided in Section \ref{sec:future}, and the conclusions are made in Section \ref{sec:conclusion}.

\begin{table*}[t]
\centering
\caption{A Comparative Analysis of AI Paradigms for Wireless Systems}
\label{tab:comparasion}
\begin{tabular}{|p{24mm}|p{40mm}|p{48mm}|p{56mm}|}
\hline
\textbf{Attribute} & \textbf{WLAM} & \textbf{Current Wireless Agentic AI} & \textbf{WELAI} \\
\hline
\textbf{Core Paradigm} 
& Data-driven pattern recognition 
& Goal-oriented planning in an abstracted space
& Embodied intelligence for physical world interaction \\
\hline
\textbf{Knowledge Base} 
& Statistical correlations from offline data 
& Policies optimized for abstract system models
& Learned physical laws and causal relationships from interaction \\
\hline
\textbf{Perception-Action Loop} 
& Open-loop: passive data input, no direct physical output 
& Logical closed-loop: virtual actions, feedback via predefined performance metrics
& Embodied closed-loop: physical actions via actuators, real-world multimodal sensory feedback \\
\hline
\textbf{Sensing Capability}
& Passive consumption of static datasets
& Reception of abstracted system states
& Active and real-time sensing using physical multimodal sensors \\
\hline
\textbf{Task Handling}
& Inference for specific, pre-defined tasks
& Decomposing goals into logical action plans
& Analyzing complex instructions and decomposing them into physically executable tasks \\
\hline
\textbf{System Role} 
& An external optimization or analytical tool
& A controller within a sandboxed digital environment
& An integrated, embodied entity that directly shapes its physical environment \\
\hline
\end{tabular}\vspace{-4mm}
\end{table*}

\vspace{-3mm}
\section{Fundamental Challenges of Current AI Paradigms in Wireless}\label{sec:limit}
As detailed in Table \ref{tab:comparasion}, this paradigm shift towards embodiment is not merely conceptual, but is fundamentally driven by the inherent characteristics of the wireless medium. While current methods have shown success in some static tasks, they fall short because they are not designed to handle the core physical realities of wireless communication. The challenges can be summarized as three key aspects:

\paragraph{Dynamic and Non-Stationary Nature} User mobility, shifting obstacles, and unpredictable interference cause channel conditions to change constantly. WLAMs that depend on static, offline-learned representations of the environment may degrade upon deployment, as their internal knowledge no longer mirrors the external world \cite{robustNetwork}.

\paragraph{Detached from the Physical World} 
Wireless communication is governed by the laws of electromagnetics and constrained by hardware, yet the AI is often treated as an external observer rather than an entity deeply embedded within the physical layer. Decisions made by disembodied agents are based on statistical knowledge from the training dataset and fundamentally limited due to a lack of awareness of real-world physical rules and conditions.

\paragraph{Lack of Action-causal Understanding}

Beyond merely adapting to the volatility of wireless environments, wireless agents face a more profound action-causal relation: its own actions actively shape this environment. While agents in many digital domains exert only limited or indirect influence on their environments, actions within a wireless system immediately and directly alter the physical state of the electromagnetic environment. Adjusting transmission power, selecting a frequency band, or steering an antenna beam would induce a direct and measurable physical effect. Models lacking embodiment with physical sensors or actuators are inherently limited in perceiving this causal relationship. 

\par
These intrinsic properties highlight a crucial insight: wireless systems are not passive data sources but active, responsive environments. Intelligence within such systems must therefore be equally active and responsive, seamlessly integrated into a continuous perception-action-learning cycle, which is the core argument for WELAI. Consequently, designing future wireless intelligence without acknowledging these intrinsic properties will lead to systems that are not robust or adaptable to real-world conditions. The paradigm must shift from pure data-centric inference to physically-grounded interaction.

\section{The Principles of WELAI Paradigm}\label{sec:principles}
Formally, WELAI is defined as a class of AI systems where intelligent agents, powered by large foundation models, directly perceive, influence, and continually learn from their physical wireless environment through a closed loop of action and sensory feedback.

\begin{figure*}[t]
	\begin{center}
		\centerline{\includegraphics[width=0.8\linewidth]{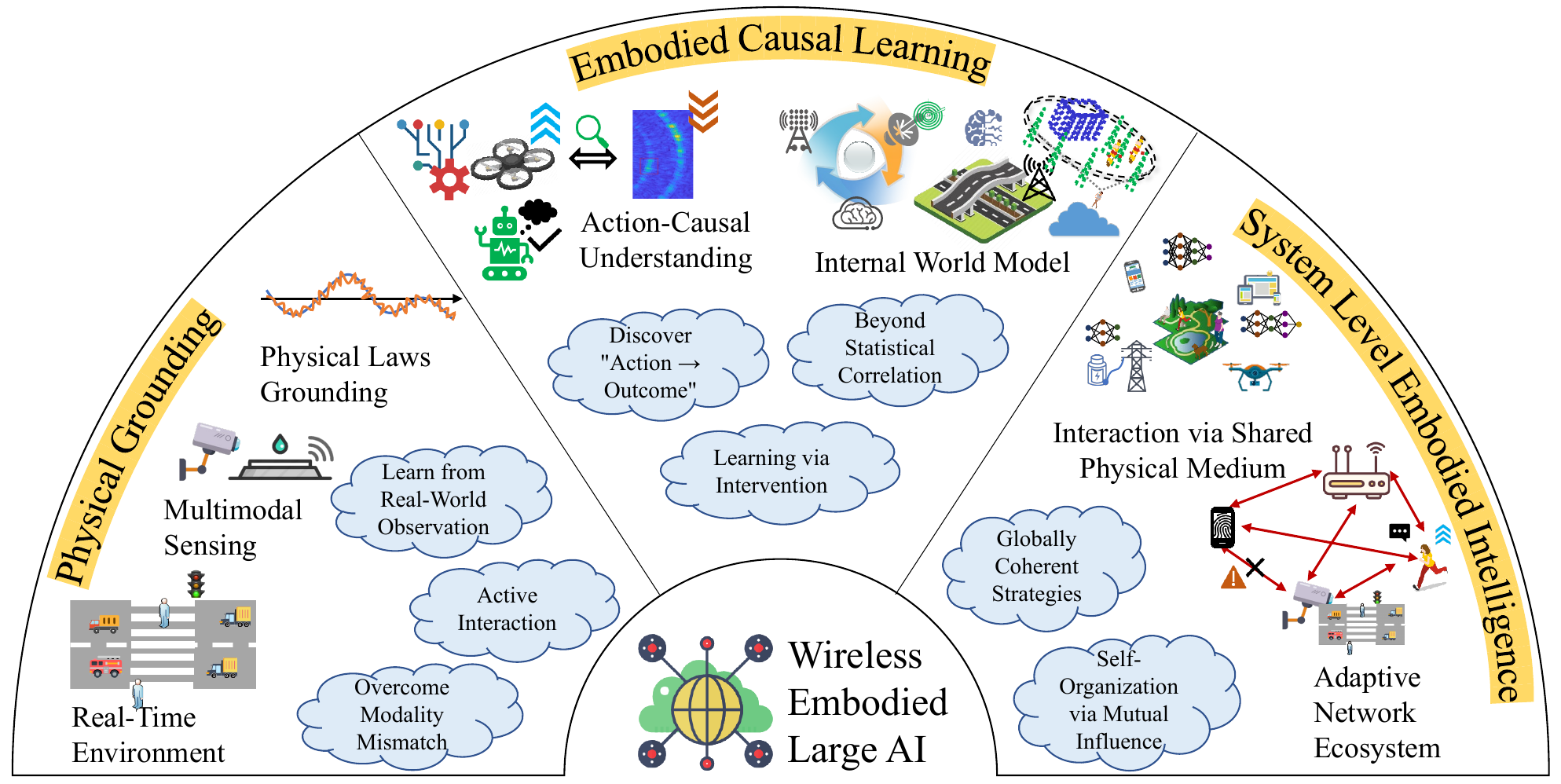}} 
		\captionsetup{font=footnotesize, name={Fig.}, labelsep=period} 
		\caption{\,  The core principles of the WELAI paradigm. 1) Physical grounding, which enables WELAI to directly links external perceptions with wireless phenomena through active physical interaction, allowing the AI to truly understand physical cause and effect. This overcomes the disconnection between knowledge and physical reality found in traditional models. 2) Embodied causal learning, which enables an agent to learn the direct impact of its own actions on network outcomes through active intervention and trial-and-error. This builds causal knowledge that goes beyond the statistical correlations in offline data. 3) System-level embodied intelligence, where embodied agents achieve complex, system-wide behaviors by interacting through a shared physical medium.}\vspace{-8mm}
		\label{fig:principle}
	\end{center}
\end{figure*}
The WELAI paradigm is built upon a set of core principles that distinguish it from traditional, disembodied AI approaches. These principles are keys that differentiate WELAI from prior paradigms. As shown in Fig. \ref{fig:principle}, these principles redefine the relationship between the AI model and its operational environment, shifting from passive data processing to active, physically grounded learning and interaction, treating the network as a holistic ecosystem rather than through isolated adjustments. They form the theoretical foundation for creating truly autonomous wireless systems that can understand, adapt to, and shape their physical surroundings.
Existing frameworks such as integrated information theory \cite{debbah2} provide ways to quantify key aspects of these principles.
The principles include:

\paragraph{Physical Grounding}
This is the foundation of intelligence. It expects all knowledge to be anchored in the physical laws governing the environment. Conventional models might recognize a truck in an image but lack any intrinsic comprehension of how the physical object will attenuate electromagnetic waves. WELAI aims to overcome this issue by pushing an agent to learn from its own continuous and embodied experience. For instance, a WELAI-enabled vehicle forges its understanding of signal blockage not from a pre-labeled dataset, but through direct touch with real world. Its light detection and ranging (LiDAR) system perceives a truck obstructing the path, and then measures a severe drop in its own received signal strength as the consequence. This event allows the agent to build a robust and explainable connection between a physical event and an electromagnetic phenomenon. This process of validation based on real-world physics ensures the internal representations are truly robust, contextually relevant, and contribute to an explainable WELAI.

\paragraph{Embodied Causal Learning} 
This defines the method of achieving intelligence. WELAI operates on embodied causal learning, advancing beyond purely correlational analysis. This principle focuses on how the agent learns to act effectively by discovering the causal links between its own actions and their outcomes, making learning an embodied process.
This also advances prior attempts to achieve causal reasoning in the wireless domain \cite{debbah2}. While causal reasoning proposes using virtual interventions to understand system dynamics, WELAI uses physical actions. Specifically, a disembodied model might learn that lower signal quality correlates with higher error rates, but a WELAI agent learns that its own experimental action of reducing transmission power causes a drop in signal quality and a subsequent rise in error rates. This allows the agent to build a robust, causal knowledge of real world, enabling it to reason effectively and predict the likely outcome of an action, rather than simply repeating patterns observed in past data. This makes up the concept of active sensing or interaction, which we define as a process where the agent itself proactively initiates a physical action to probe the environment and measure the resulting response. The sensory feedback provides direct evidence for the agent to build and continuously validate its internal causal world model.

\paragraph{System Level Embodied Intelligence}

This principle defines the ultimate form of intelligence, facilitating a paradigm shift from fragile, centralized network control to robust, system-wide intelligence achieved through embodied emergence \cite{wei2022emergent}. It posits that globally coherent strategies can arise from decentralized agents interacting implicitly through the shared physical medium, where the wireless environment itself becomes the communication substrate. This coordination is not random but is achieved as each agent learns an internal causal world model that includes the actions of other agents as influential variables.
For instance, when a base station agent perceives interference from an unmanned aerial vehicle (UAV), it does not merely react to the signal degradation. Instead, it perceives this change and updates its causal model to understand the interference as an effect of the UAV. It can then employ methods like game-theoretic reasoning \cite{debbah2} to infer the objectives and the future behavior of the UAV, leading to proactive co-adaptation like adjusting its own reception patterns.
Through this process, agents learn to anticipate how their actions will influence others, which is the key to the emergence of implicit coordination strategies like turn-taking or spontaneous resource sharing. Consequently, complex system-level behaviors, such as interference coordination and dynamic load balancing, emerge from these sophisticated local interactions, enabling a truly scalable, resilient, and adaptive network ecosystem without explicit programming or a central server.

\section{WELAI System Structure}\label{sec:design}
\begin{figure*}[t]
	\begin{center}
		\centerline{\includegraphics[width=0.87\linewidth]{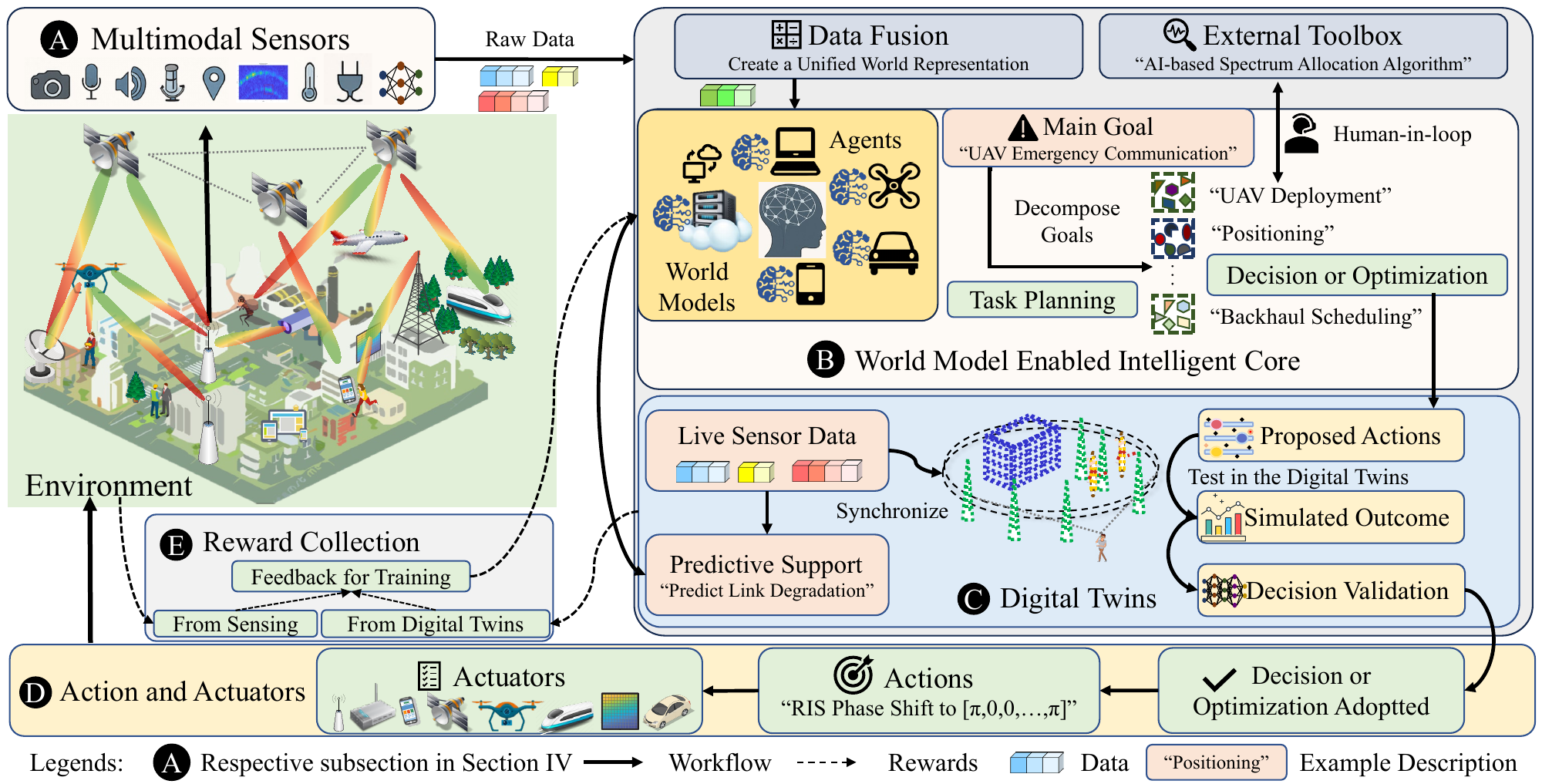}} \vspace{-2mm}
		\captionsetup{font=footnotesize, name={Fig.}, labelsep=period} 
		\caption{\, The system architecture and operational workflow of the WELAI system. (A) Multimodal sensors gather raw data from the environment, which is fused to create a unified world representation for the intelligent core. (B) The world model enabled intelligent core, guided by a main goal, performs task planning and makes decisions. It can consult external tools or a human-in-the-loop. (C) Proposed actions can be pre-validated for safety and efficacy in a digital twin, which is synchronized with live sensor data. (D) Validated decisions are translated into physical actions by actuators, which directly influence the environment. (E) The environmental changes are captured as rewards, which are used to train and refine the intelligent core, closing the learning loop.}\vspace{-9mm}
		\label{fig:welai}
	\end{center}
\end{figure*}

To implement the principles of the WELAI paradigm, a specialized system architecture design based on the three WELAI principles is required. 
Specifically, physical grounding is mainly realized through advanced multimodal sensors, physical actuators and the intelligent core. Embodied causal learning is powered by the intelligent core that learns from interaction, often accelerated and de-risked by a digital twin. Finally, system-level embodied intelligence emerges by interacting implicitly through the shared physical medium and is enabled by the overall framework.
This section details the functional components of a WELAI system, which work in concert to bridge the gap between abstract intelligence and physical interaction, as illustrated in Fig. \ref{fig:welai}. 

\subsection{Multimodal Sensing}
The gateway for WELAI to perceive the physical world is its multimodal sensing capability.
This marks a significant difference from traditional systems that rely almost exclusively on radiofrequency (RF)-centric measurements.
Its primary function is to capture a rich, real-time tapestry of data spanning multiple modalities, including both RF information and data from visual, auditory, radar, LiDAR, temperature, humidity, location sources, and even service demand from application-layer data. 
For instance, in a vehicular network, LiDAR can perceive physical obstructions while RF sensors measure the resulting signal blockage, allowing the agent to ground RF phenomena in physical objects.
To effectively support a physically grounded WELAI system, this module must meet several key requirements.
It demands low-latency data acquisition to handle fast-changing dynamics, robustness against adverse weather, interference or sensor failure.
Efficient and robust data fusion capabilities remains a significant bottleneck, crucial to create a unified world representation from heterogeneous and noisy data \cite{xuLMMs}.

\subsection{World Model Enabled Intelligent Core}
A large-scale world model \cite{matsuo2022deep} serves as the intelligent core of the WELAI architecture, functioning as a tightly coupled cognition-execution engine. This core processes fused data from the sensor module to build, use, and maintain a predictive representation of the world, enabling a continuous cycle of reasoning and action.
The world model learns representation and the regular pattern of the environment dynamics directly from interaction. Its learning process is uniquely suited for WELAI. It is not trained solely on task-specific rewards, but rests predominantly on a task-agnostic, unsupervised objective to reconstruct rich sensory inputs. This pushes a WELAI agent to form a robust understanding of the underlying structure and laws of the world, allowing the agent to perform physically and causally grounded what-if analyses \cite{hafner2025mastering,matsuo2022deep}. This world knowledge then enables the agent to select an efficient path to a goal by consequence prediction via such what-if analyses.
It also helps the world model understand the meaning of an action in the context of the physical environment, allowing complex behaviors to emerge from simple commands without exhaustive instructions \cite{debbah1}.
\par
The world model enabled core underpins multiple functions by providing intelligence on demand, including:
\subsubsection{Task Planning}
The task planning module decomposes high-level goals into a sequence of actionable intermediate objectives. It leverages predictions and reasoning from the intelligent core to enable proactive, goal-driven behavior. For example, ensuring emergency communication may involve subtasks such as UAV deployment, dynamic positioning, and intelligent backhaul scheduling.
This module supports long-term decision-making by dynamically redrafting plans when the intelligent core updates its understanding, when objectives shift, or when subtasks fail. Besides, in multi-agent scenarios, such as coordinated reconfigurable intelligent surfaces (RIS) networks or UAV fleets, it also handles inter-agent coordination and conflict resolution. Planning can be performed directly by the intelligent core or integrated with external knowledge sources or human-in-loop intervention mechanism.

\subsubsection{Decision or Optimization}
This module translates task plans and situational context into actionable control commands executable in physical environments. In wireless communication systems, typical decisions encompass adaptive parameter tuning, spectrum allocation, and RIS configurations aimed at optimizing multi-user network performance.
Depending on the complexity and requirements of specific tasks, the module dynamically selects appropriate decision or optimization tools. These tools may be the direct output of the intelligent core, or may be external AI-based or conventional optimization solvers or heuristic-based methods. Robustness against uncertainties and unexpected variations is improved by integrating constraint management layer and fallback strategies, and pre-validating through digital twins prior to physical actuation.
\par
Building such an intelligent core is technically demanding, as it must learn from high dimensional, noisy data and support long-time continual learning without catastrophic forgetting. Safety mechanisms like backup systems must be in place when prediction confidence is low or when decisions may lead to instability, such as oscillations or service disruptions. Transparency and interpretability are also essential for trustworthiness.

\subsection{Digital Twins}\label{sec:digitaltwin}
The digital twin is a core enabler of WELAI, functioning as a high fidelity virtual replica of the wireless physical environment \cite{DigitalTwin, taoDT}. In WELAI systems, it serves several critical roles. Primarily, it provides a safe and efficient virtual environment where the agent can accelerate the training of its internal world model and validate new policies before real-world deployment. This allows agents to rapidly test thousands of actions to learn complex strategies without real-world risk. Additionally, the twin offers predictive support for real-time control by forecasting near future conditions based on live data, such as link degradation due to mobility. This foresight allows the WELAI agent to execute proactive actions.

However, realizing an effective digital twin presents significant challenges. The required fidelity of the twin is task-dependent and achieving it can be complex. More critically, maintaining synchronization between the physical world and the digital replica is a major hurdle, as communication delays can degrade the accuracy of the twin, especially in highly dynamic environments. One promising direction is the use of decentralized, edge-enabled architectures to minimize these delays and provides a common situational awareness space that facilitates collaborative planning and coordination. Furthermore, the system must account for inevitable mismatches in the twin model. Therefore, the digital twin must incorporate self-calibration mechanisms using sensory data.

\subsection{Action and Actuators}
The decisions or optimization outcomes are subsequently transformed into actions. Actuators are the physical interface translating decisions into actions. They receive digital control signals and implement changes in the wireless or physical environment. 
Examples include RIS, which manipulate wave propagation, UAVs that dynamically adjust position, and software-defined radios (SDR) platforms with programmable communication parameters. The actuators complete the perception-reasoning-action link, enabling WELAI to actively shape its environment.
Key design requirements for actuators include rapid and precise execution, broad and finely adjustable parameter ranges, operational reliability under diverse conditions, and energy efficiency. Advanced actuator functionalities may also necessitate supplementary technologies such as self-calibration mechanisms, synchronization protocols, and coordination with the intelligent core.

\subsection{Environmental Impact and Reward Collection}
The actions of actuators complete the perception-action loop, allowing a WELAI agent to strategically shape its wireless environment toward objectives such as enhanced coverage, reduced congestion, and stable connectivity. To guide this process, the system continuously evaluates the environmental impact of these actions through reward signals. These signals quantify the effectiveness of a given strategy in achieving system goals and are fundamental to adaptive learning. However, designing effective reward functions presents challenges, especially in complex scenarios involving delayed feedback or multi-agent settings, which can complicate credit assignment and hinder stable learning. To enrich the learning data, feedback is derived not only from physical interactions but also from high-fidelity simulations within the digital twin. This combined stream of reward-driven data enables the intelligent core to apply advanced reinforcement learning (RL)-based methods, facilitating embodied causal learning.

\section{Case Study}\label{sec:case}
To illustrate the WELAI paradigm, this section presents an application in intelligent vehicular networks \cite{v2v}. The case demonstrates how a vehicle, acting as a WELAI agent, can leverage embodied intelligence to enhance its perception and decision-making. While focused on a vehicular system, this case validates the fundamental WELAI design principles and shows its broader potential applicability to other next-generation wireless systems.

\subsection{Task Description}
The task involves a WELAI agent vehicle navigating complex urban environments. The primary goal is to operate safely and efficiently by overcoming key challenges inherent to the internet of vehicles (IoV), such as the low correlation of multimodal data, high uncertainty in the driving environment, and insufficient inherent intelligence. The vehicle is equipped with multimodal sensors (e.g., cameras, LiDAR) and can use vehicle-to-vehicle (V2V) communication. System performance is evaluated on a range of metrics, including route completion percentage, driving scores, vehicle stability, and efficiency.

\subsection{Proposed WELAI-based Solution}
The proposed solution is architected in three main units.

The process begins in the multimodal sensing unit. This unit first uses a residual network to extract features from camera images and LiDAR-based bird-eye view data. It then employs cross-blocks to recover missing data and preserve information, and a multi-head self-attention mechanism to integrate the complementary image and LiDAR modalities. Concurrently, V2V communication data, containing the position and speed of nearby vehicles, is processed by a dense neural network. Finally, a transformer fuses all three modalities to produce a single, comprehensive description of the driving environment.

This fused representation is then passed to the intelligent decision-making unit, which functions as the WELAI intelligent core. A multimodal LLM (MLLM) interprets prompts that describe the current environmental state and the high-level driving intention. Based on its analysis, the MLLM generates optimal driving decisions and continuously refines them. A critical feature is the continuous-learning mechanism: the outcome is assessed after an action is executed, and the MLLM engages in self-reflection based on this feedback, storing the results to improve future performance. This enables the agent to balance long-term planning with real-time responses to unexpected events.
This architecture creates a powerful perception-action loop central to the WELAI concept. The actions of the agent influence its own subsequent perceptions and decisions. A physical action, like a lane change, immediately alters the driving environment, and this change is captured by the sensors, creating a new prompt for the MLLM. 

Furthermore, through the actuators in the communication optimization unit, the agent can take actions that directly shape the surrounding wireless environment. The resulting impact on communication quality serves as another form of feedback, allowing the agent to learn the consequences of not just its physical actions, but its communication strategies as well.

\begin{figure}[t]
	\begin{center}
		\centerline{\includegraphics[width=\linewidth]{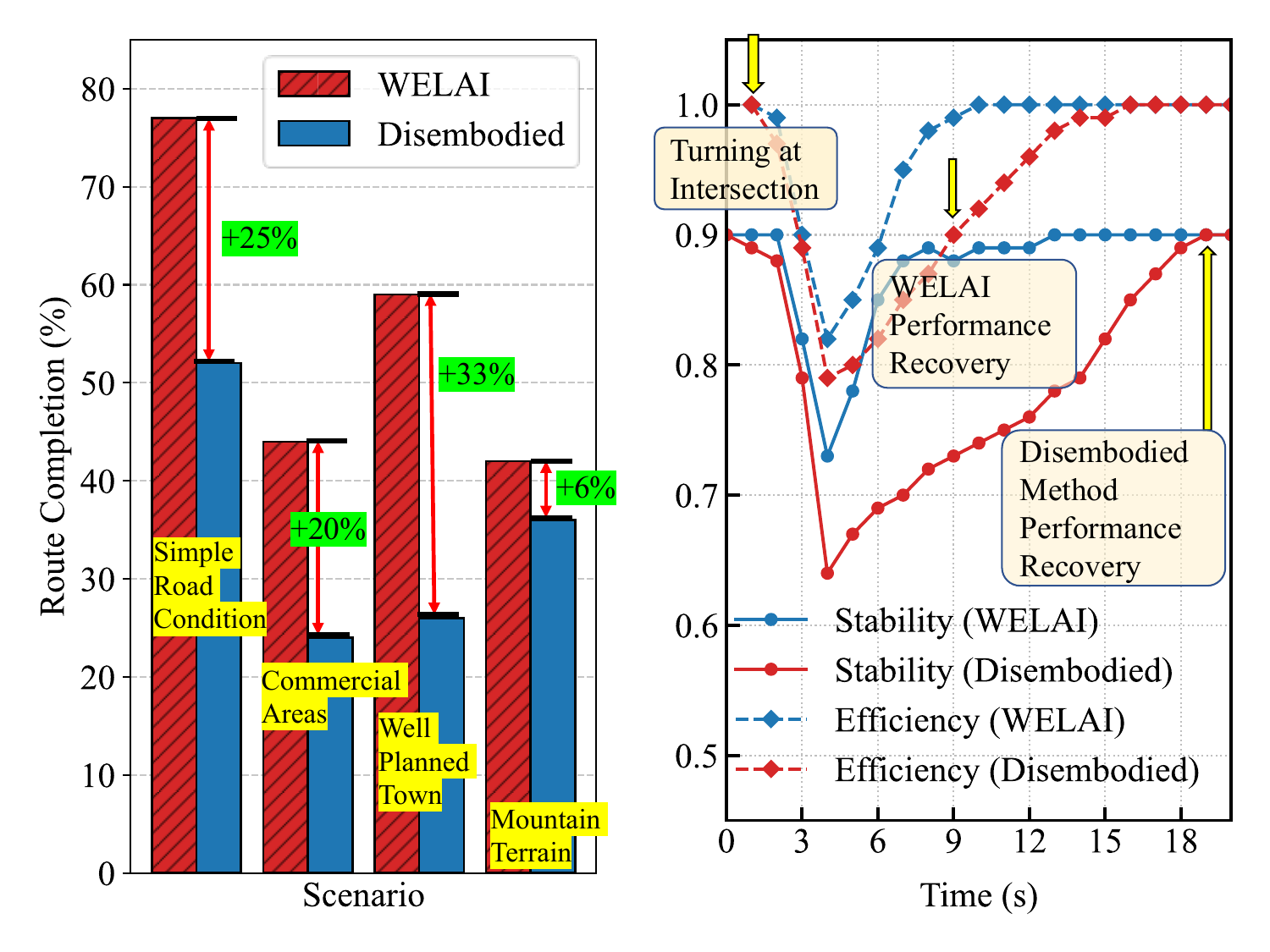}}\vspace{-4mm}
		\captionsetup{font=footnotesize, name={Fig.}, labelsep=period} 
		\caption{\, A comparison between the WELAI agent and baseline, across 4 scenarios and during a turn at an intersection.}\vspace{-10mm}
		\label{fig:case}
	\end{center}
\end{figure}
\subsection{Evaluation Results}
The system was evaluated in various long-route scenarios to compare the proposed method against the TransFuser baseline, a conventional disembodied end-to-end model that uses transformers to fuse data for imitation learning-based waypoint prediction.
In long-route evaluations, the WELAI agent demonstrated superior performance, shown in Fig. \ref{fig:case}. For instance, in a well-planned town, the route completion of the WELAI agent was approximately 59\%, more than double the 26\% achieved by the baseline. This significant performance advantage held across all tested environments, including simple roads and mountain terrain. To further quantify the benefits of embodiment, additional experiments measured adaptability using vehicle stability (comfort) and driving efficiency metrics. In an experiment of turning maneuver in an intersection, the WELAI agent showed greater resilience, with its stability score dropping less than the baseline system. The agent also adapted significantly faster, recovering full performance in approximately 9 seconds, which was twice as fast as the disembodied baseline.

\subsection{Discussion and Generalization to Broader Applications}
The performance gains are a direct result of the WELAI paradigm. 
First, the multimodal fusion unit actualizes physical grounding, creating a robust environmental understanding that outperforms systems with fewer sensing modalities. Second, the MLLM-powered intelligent core realizes embodied causal learning with its continuous-learning and reward mechanism. It moves beyond simple pattern recognition by learning from the direct consequences of its actions, enabling more reliable decisions in uncertain scenarios. Finally, the dynamic communication adjustment demonstrates system-level embodied intelligence, where the agent actively shapes its network interactions for system-wide benefit.

The effectiveness of the WELAI-based method in this case validates its potential for broad application. While the presented case focuses on the core perception-action loop, its success underscores the potential of the complete WELAI architecture, including components like digital twins and external tool use for even more complex scenarios. 
The core mechanisms underlying WELAI are broadly applicable across other wireless domains, especially sensor-rich environments.
For example, in dynamic factory environments, WELAI agents can interpret data from a multitude of sensors to adaptively coordinate spectrum usage and physical movements among mobile robots, enhancing production efficiency and safety. In UAV-based relay networks, WELAI systems can anticipate channel disruptions due to movement or obstruction and proactively adjust flight paths and beam patterns.
Even in infrastructure-less emergency scenarios, embodied agents can leverage localized sensing to self-organize and form ad-hoc communication networks without reliance on centralized coordination, accelerating disaster recovery efforts.

\section{Future Research Directions}\label{sec:future}
The realization of WELAI is contingent upon advancements across several key technological frontiers. We outline 5 pivotal research directions, progressing from the physical interface with the environment to the core intelligence models, and ultimately to the system-level protocols and standardization.

\subsection{Advanced Physical Perception and Interaction}
Effective embodiment begins with a fine-grained, low-latency interaction with the wireless environment, which necessitates a co-evolution of perception and actuation technologies. On the perception front, future research must push beyond traditional channel state information to build a richer environmental understanding. While advanced paradigms like integrated sensing and communication can map the physical environment using the communication signals themselves, such methods often operate in isolation from the intelligent core. Therefore, the central challenge is not merely data acquisition, but developing lightweight algorithms that can extract rich, real-time information and advanced fusion techniques to integrate these diverse multimodal signals into a single, coherent representation for the agent.
On the actuation side, the focus must be on creating a responsive physical interface that can translate agentic intent into precise physical layer behavior. Real-time controllable hardware, including digital beamformers, next-generation RIS, and fully programmable waveform generators, are the primary enablers. A key research goal is the reduction of control latency to the sub-millisecond level, allowing for a tightly integrated perception-action loop where sensing informs action, and action purposefully enhances subsequent sensing, all within the strict time constraints of wireless communication.

\subsection{Specialized Model Architectures for Edge Embodied Intelligence}
To achieve WELAI, a fundamental limitation of existing LLMs is that their knowledge is purely statistical, derived from data patterns rather than being grounded in the physical laws of the real world. A primary research direction is thus the creation of multimodal wireless foundation models that address the unique causal nature of wireless multimodality. 
However, deploying such large models presents significant computational and energy constraints, particularly at the wireless edge. Addressing this requires a focus on efficiency. Promising strategies include advanced model compression and the adoption of more efficient architectures, such as state-space models like Mamba, or hybrid architectures like reusable weight and key value. While promising, further research is needed to better integrate their capabilities for embodiment.

\subsection{Language-Driven Control and Human-AI Interaction}
Another critical direction is the development of language-driven control mechanisms to facilitate advanced human AI collaboration. This requires agents that can deeply understand complex, and potentially ambiguous human instructions through interactive dialogue, aiming for a language-driven wireless system capable of sophisticated self adjustment and self organization. Furthermore, to support the long horizon decision making required for tracking environmental non stationarities, specialized memory architectures are vital for processing long and irregularly sampled sequences from the physical world.
For this, adaptive methods like test-time training (TTT) present a compelling research avenue. By allowing the model to dynamically update its weights during inference, TTT could enable an agent to specialize to a specific human intent and conversational nuances online.

\subsection{External Tools Use and Multi-Agent Coordination}
To enhance operational efficiency, WELAI agent may not be a monolithic entity attempting to solve every problem internally. Future research should focus on frameworks that allow the agent to utilize existing and validated algorithms, such as beamforming optimizers, as external tools. This requires establishing a comprehensive tool library and enabling the AI core to autonomously format data to meet the specific application programming interface (API) requirements for invoking these tools. This approach leverages the best of both learned models and traditional, validated algorithms.
In addition to interacting with tools, in multi-agent contexts, agents must also coordinate with each other. The development of standardized and efficient multi-agent communication protocols is essential. These protocols will enable WELAI agents to negotiate, share situational awareness, and deconflict actions to achieve emergent, system-level objectives without explicit central control.

\subsection{Pathways to Standardization}
Standardization for WELAI is not merely an industrial process but a distinct research challenge, centered on resolving the fundamental tension between the flexibility required for AI-driven adaptation and the robustness needed for interoperable network operation. The goal is to formulate standards that strike a balance between ensuring operational safety and enabling adaptive, emergent intelligence within WELAI systems. Key research questions in this area include: 1) How to design dynamic APIs for sensors and actuators that can support the exploratory nature of learning agents while guaranteeing operational safety and predictability. 2) How to define intelligent guardrails instead of static rules. This involves creating protocols that can certify the behavior of an evolving agent in real time, preventing network instability without stifling beneficial adaptation. 3) How to formulate coordination protocols for multi-agent systems where agents are adaptive, potentially opaque black boxes. Research is needed to ensure that their interactions lead to globally stable and efficient outcomes.

\section{Conclusion}\label{sec:conclusion}
In this paper, the WELAI paradigm was introduced to bridge the gap between the cognitive capabilities of AI and the physical dynamics of wireless networks. The limitations of current disembodied AI models were first identified, followed by an exploration of the core design principles and system structure of WELAI, where the shift from passive observation to active embodiment was emphasized. The potential applicability and advantages of the WELAI paradigm were illustrated through prospective application scenarios and a detailed case study. To guide future work, key research directions and enabling technologies essential for the realization of this paradigm were outlined. The adoption of the proposed embodiment principles is considered essential for engineering the adaptive, robust, and autonomous wireless systems of the future.

\bibliographystyle{IEEEtran}
\bibliography{embodied}

\end{document}